\documentclass[a4paper,11pt]{article}

\usepackage{contribution}

\newcommand{\weblink}[2][]{%
    \ifthenelse{\equal{#1}{}}%
    {\textnormal{\url{#2}}}%
    {\textnormal{\href{#2}{#1}}}%
}

\newcommand{\acknowledgements}[1]{%
  \bigskip\bigskip
  \textsf{\textbf{\Large Acknowledgements}} \\[2ex]
  {#1}
  \bigskip
}

\def\beq{\begin{equation}}
\def\eeq#1{\label{#1}\end{equation}}
\def\eeqn{\end{equation}}

\def\beqa{\begin{eqnarray}}
\def\eeqa#1{\label{#1}\end{eqnarray}}
\def\eeqan{\end{eqnarray}}

\let\bar=\overbar

\def\etal{{\it et al.}}

\def\Dslash{\not{\hbox{\kern-4pt $D$}}}
\def\dslash{\not{\hbox{\kern-2pt $\del$}}}

\def\msb{{\bar{\ssstyle M \kern -1pt S}}}

\newcommand{\contribution}[7][]{%
  \clearpage
  \thispagestyle{plain}
  \ifthenelse{\equal{#1}{}}
  {\hypersetup{pdftitle={#2}}}
  {\hypersetup{pdftitle={#1}}}
  \hypersetup{pdfauthor={{#3} {#4}}}
  {\centering\normalfont\LARGE\bfseries\sffamily #2 \par\nobreak}
  \lhead{}
  \chead{%
    \textit{\footnotesize XIV International Conference on Hadron Spectroscopy
      (\weblink[\textit{hadron2011}]{http://www.hadron2011.de}), 13-17 June 2011, Munich, Germany}%
  }
  \rhead{}
  \bigskip
  \begin{center}
    {#3} {#4}\ifthenelse{\equal{#6}{}}{}{\footnote{\weblink[#6]{mailto:#6}}}
    \ifthenelse{\equal{#7}{}}{}{#7} \\
    \textit{#5}
  \end{center}
  \bigskip
}

\renewcommand{\abstract}[1]{%
  \begin{center}
    \begin{minipage}{0.85\textwidth}
      \begin{footnotesize}
        #1
      \end{footnotesize}
    \end{minipage}
  \end{center}
  \bigskip
}

\begin{document}

%
%
%
%
%
{  

%

\contribution[]  
{Photoproduction of $\eta$-Mesons off $^3$He}  
{Lilian}{Witthauer}  
{Department of Physics \\ University of Basel \\
  CH-4056 Basel, SWITZERLAND }  
{lilian.witthauer@unibas.ch}  
{for the A2 Collaboration}  
\abstract{
Photoproduction of $\eta$-mesons off $^3$He has been studied at the MAMI 
accelerator using the Crystal Ball/TAPS detector setup. 
The total cross section for the coherent $\eta$-photoproduction was measured with improved
statistical quality. 
Both, the total and differential cross sections show evidence for dominant final state
interaction.
Additionally, the photoproduction of $\eta$-mesons off quasi-free protons and neutrons was
studied. The
preliminary cross section on the neutron confirms the narrow bump-like structure at $W
\simeq 1.7$ GeV, which was already seen in different experiments on the deuterium target
\cite{Graal, Jaegle, Jaegle2, Domi, LNS}.}
\section{Introduction}
Photoproduction of mesons is an ideal tool to investigate the meson-nucleon (-nucleus) interactions.
An important question is whether the properties of the strong interaction allow the
formation of meson-nucleus bound states. The best candidate for such a bound state is the
$\eta$-meson. Already in the 1980s Bhalerao, Liu and Haider \cite{Liu,Haider} found that an
attractive $\eta$N s-wave interaction might lead to the formation of quasi-bound $\eta$-nucleus
states, the so-called $\eta$-mesic nuclei. Such quasi-bound states should give rise to an
enhancement at
the threshold of the cross section relative to the expectation for
phase space behavior. Such threshold behaviours have been previously studied in hadron and photon
induced reactions for $\eta$-$^{3}$He \cite{Pfeiffer, Mersmann} and $\eta$-$^{4}$He
\cite{Hejny1,Hejny2} systems. With the $4\pi$-detector in Mainz, this experiment was able to improve
the statistical quality of the coherent $\eta$-photoproduction off $^3$He drastically. The results
will be discussed below.\newline
Besides the investigation of $\eta$-mesic nuclei, this experiment was used to study the excitation
spectrum of the nucleon. 
In particular, total cross sections of $\eta$-photoproduction off quasi-free protons and neutrons
have been measured. Previous experiments \cite{Graal, Jaegle, Jaegle2, Domi, LNS} have
reported a narrow structure at $W \simeq 1.7$ GeV with a width of $\sim25$ MeV in the cross section
on the neutron which is not visible for the proton.\newline
The experiment was carried out at the MAMI acceleration facility in Mainz. 
A circularly polarised tagged photon beam with energies up to
1.4 GeV and a typical energy resolution of 4 MeV was used. For a geometrical acceptance close to
$4\pi$
steradian several detectors are needed. The Crystal Ball detector (CB) surrounds the
target and is made of 720 NaI crystals. Inside the CB the Particle Identification Detector (PID) is
placed. The PID is made of 24 plastic scintillators and is used to identify charged particles. The
opening angle to the forward direction of the CB is covered using the photon spectrometer
TAPS. The TAPS detector is made of BaF$_2$ and PbWO$_{4}$ crystals and is placed $1.475$ m
in front of the target. A plastic veto is mounted in front of every BaF$_2$
crystal. The cryogenic $^3$He target is centered in CB and has a length of $5.3$ cm and a
density
of $0.069\text{ g}/\text{cm}^3$.
\section{Results}
To investigate the formation of a $\eta$-nucleus bound states, the cross
section of the coherent $\eta$-photoproduction was measured. An invariant mass analysis for each
energy and meson center-of-mass polar angle has been performed to identify the $\eta \to 2\gamma$
and $\eta \to 6 \gamma$ decay. Due to the overdetermined kinematics the missing energy was used
to separate the coherent from breakup reactions. 
\begin{figure}[htb]
  \begin{center}
    \includegraphics[width=0.329\textwidth]{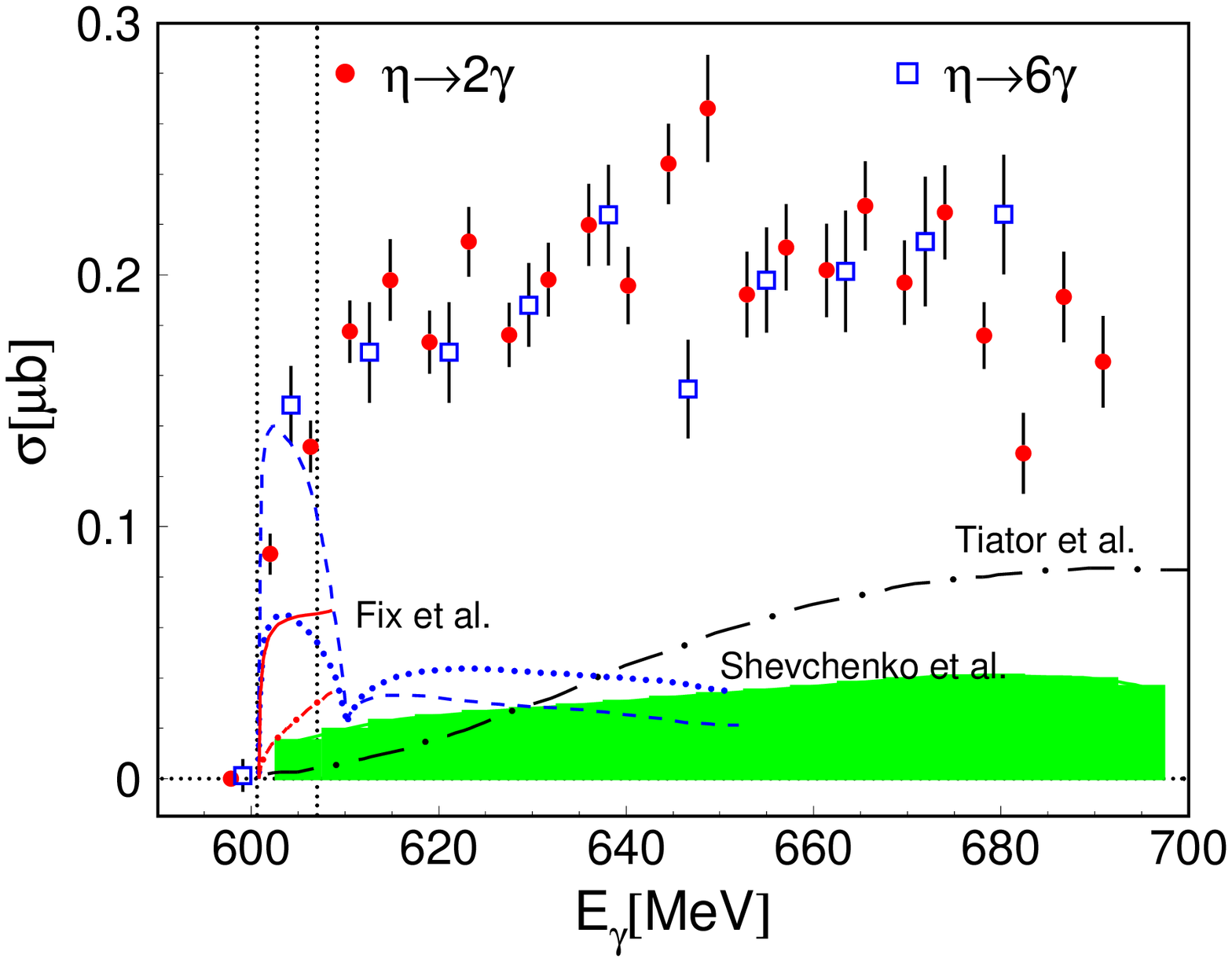}	
    \includegraphics[width=0.3\textwidth]{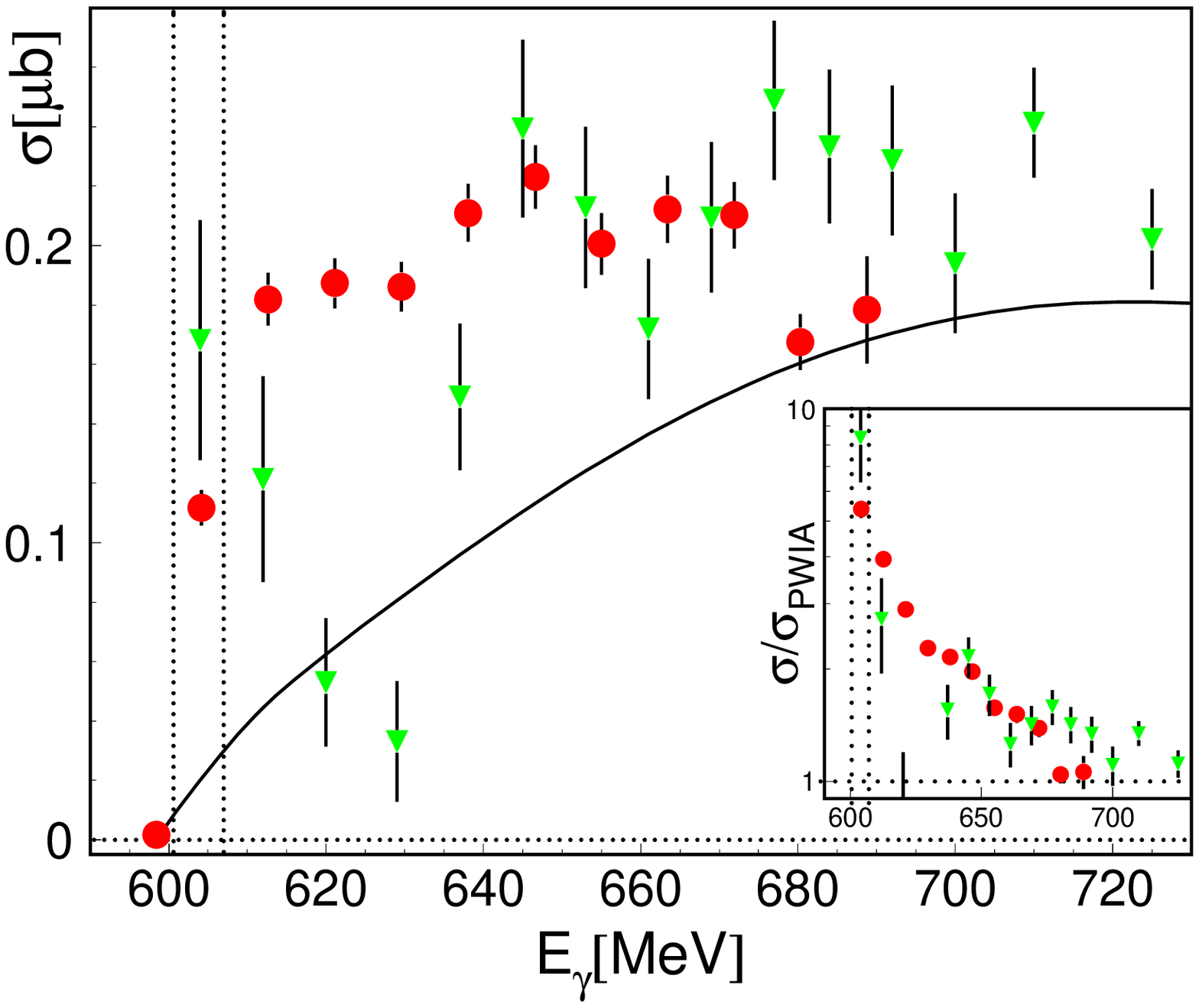}
    \caption{Left: Coherent $\eta$ cross section for $\eta \to 2\gamma$ (red) and $\eta \to 6\gamma$
(blue). Several models are indicated. Right: The average of  $\eta \to 2\gamma$ and $\eta \to
6\gamma$ is compared to a previous measurement by M. Pfeiffer et al. \cite{Pfeiffer}.}
    \label{fig:etamesic}
  \end{center}
\end{figure}
The total cross section of the two decay channels are shown in
Fig.\ref{fig:etamesic} (left). Both results are in good agreement and show a steep increase
between the coherent and the breakup threshold. All indicated models do not reproduce the data. On
the right side the average of the two cross sections is compared to an earlier
experiment \cite{Pfeiffer}. The two measurements are in agreement if one takes into account the
lower statistical quality of the older results.\newline
The $\eta$-photoproduction off quasi-free protons and neutrons has been identified with an invariant
mass analysis. The competing background, which mainly comes from $\eta\pi^{0}$ reactions, was
eliminated by cutting on the missing mass and on the coplanarity of the $\eta$-nucleon pair.
Additionally, the background was reduced by coincidence cuts and a random background subtraction.
Monte-Carlo simulations with Geant4 were used for the angle and energy dependent detection
efficiency correction. 
\begin{figure}[htb]
  \begin{center}
    \includegraphics[width=0.3\textwidth]{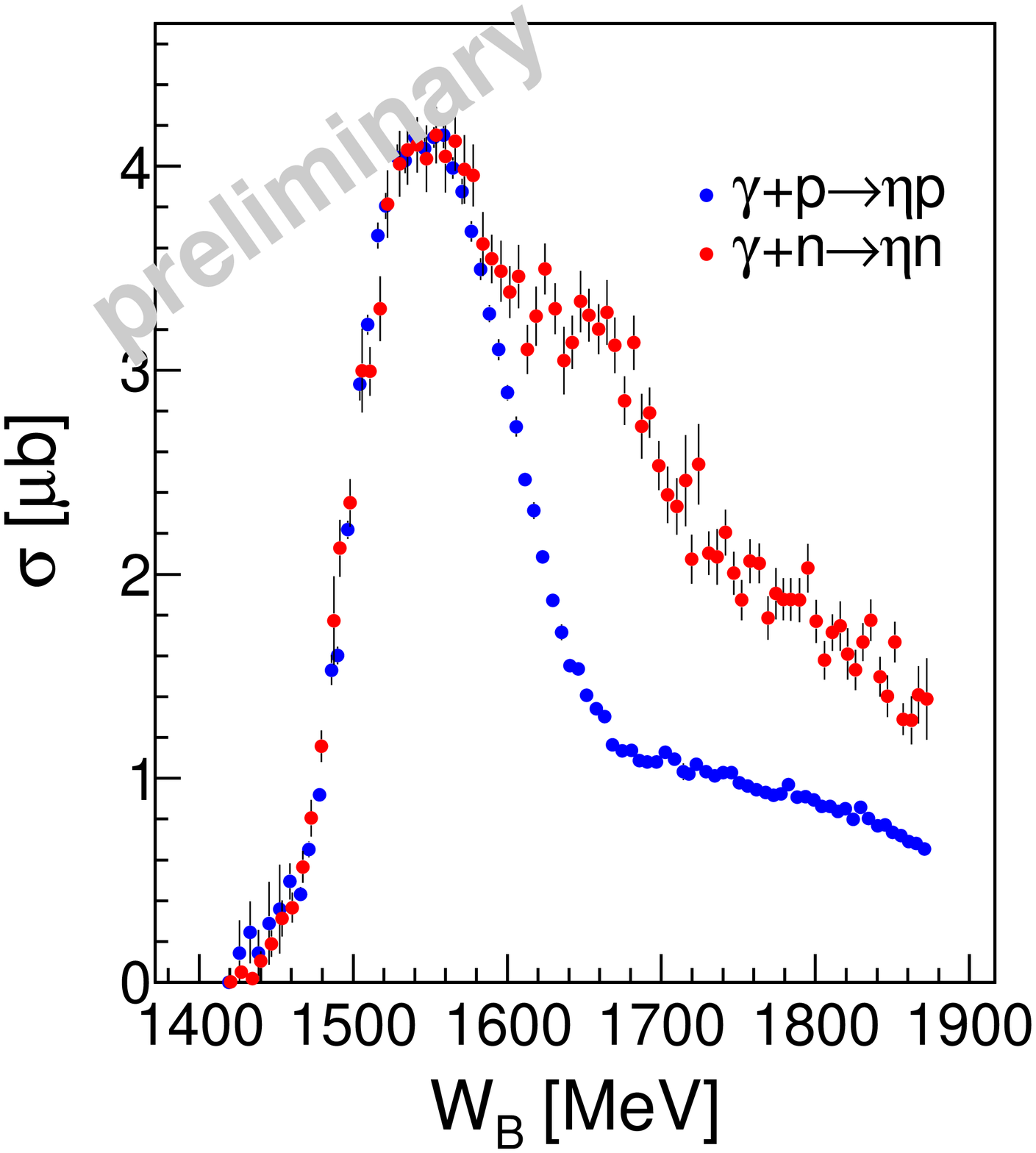}	
    \includegraphics[width=0.3\textwidth]{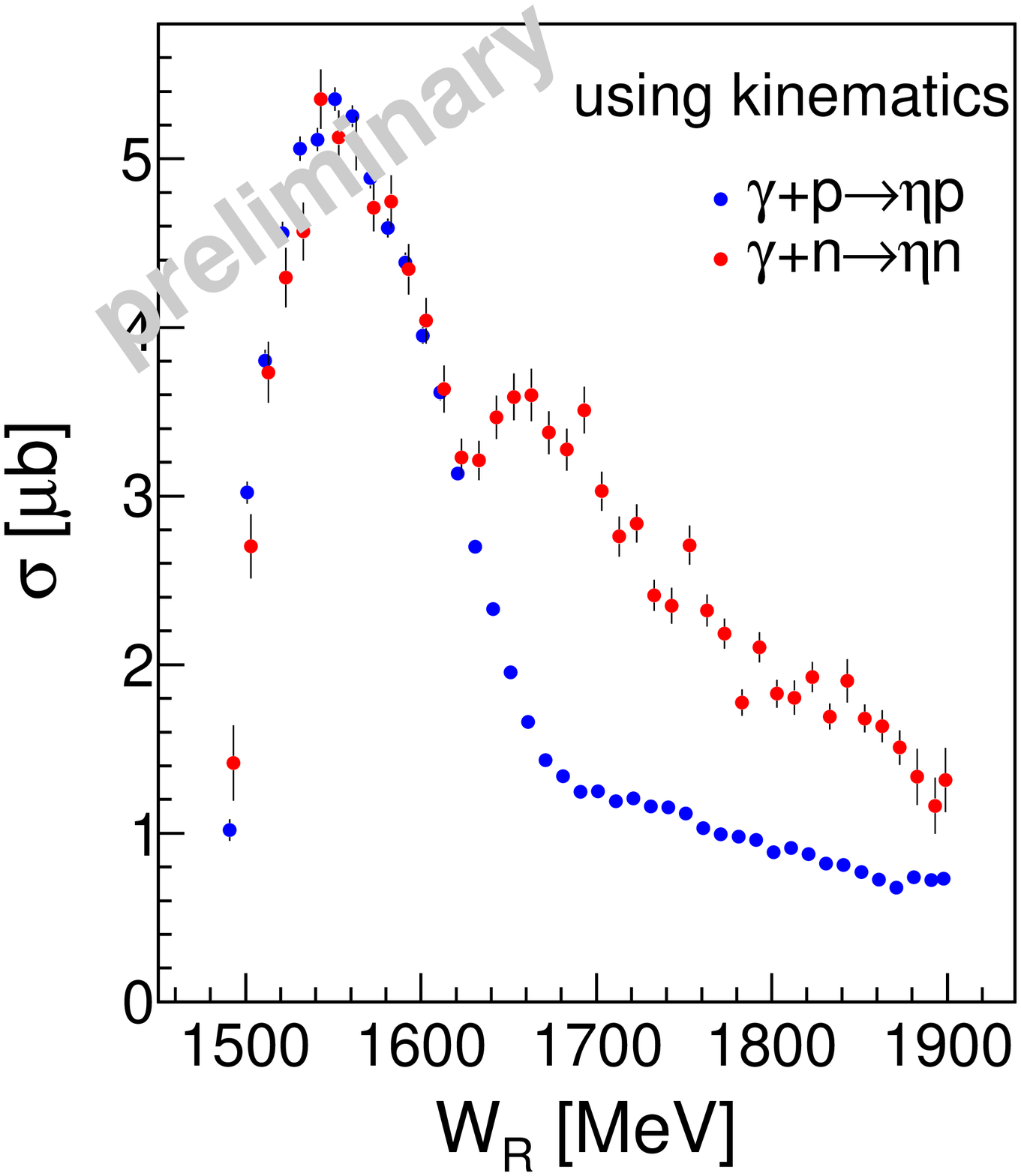}
    \includegraphics[width=0.3\textwidth]{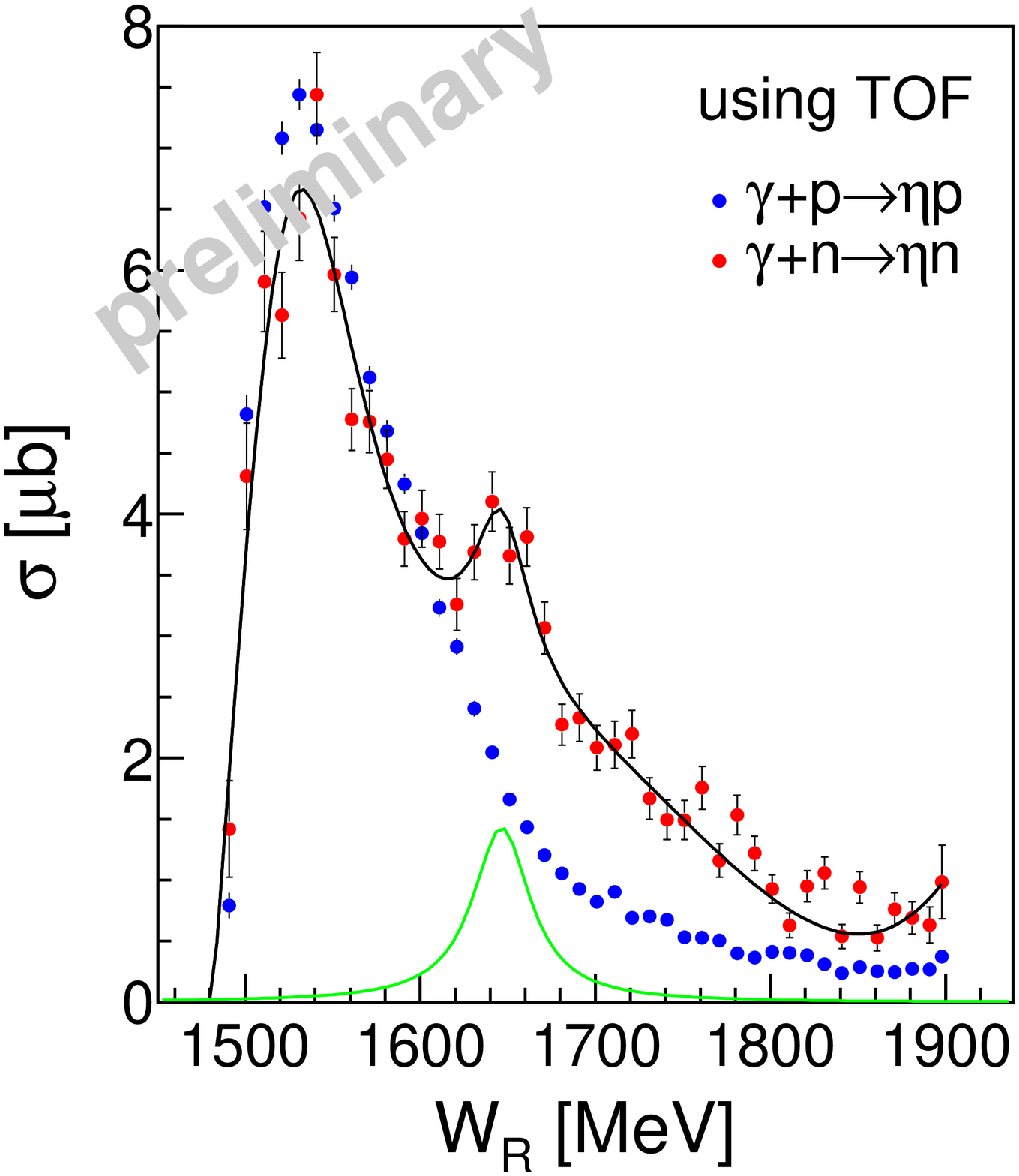}
    \caption{Total cross sections of quasi-free $\eta$-photoproduction on the proton (blue) and on
the neutron (red). The proton cross section is scaled to the neutron. The cross section on the
right-hand side is fitted with two Breit-Wigner functions and a background polynomial. The fit
yields a
width of 45 MeV for the structure, which is comparable to the experimental resolution.}
    \label{fig:qfeta}
  \end{center}
\end{figure}
On the one hand the cross section was calculated as a function of the
center of mass energy with the initial state particles:
$$ W_{B}^{2} = (P_{\gamma} + P_{N,i})^{2} = 2E_{\gamma}m_{N} + m_{N}^{2} $$
Since the momentum of the initial state nucleon is not exactly known, the structures are smeared out
due to the Fermi motion. 
On the other hand the center-of-mass energy has been calculated with the final state particles:
$$ W_{R}^{2} = (P_{\eta} + P_{N,f})^{2} $$
In this case no effects of Fermi motion are visible but the experimental resultion of
the recoil nucleon is the limiting factor.
The resulting cross sections as a function of $W_{B}$ (left) and $W_{R}$ (middle) are
visible in Fig.\ref{fig:qfeta}. In the central picture the structure around $W\simeq 1675$ MeV is
quite
narrow, whereas the structure on the left side is broadened by Fermi motion. The position of this
structure is consistent with the deuterium data by I. Jaegl\'e et al. \cite{Jaegle, Jaegle2} but is
somewhat
broader. This is caused by the fact that the kinematical reconstruction of the recoil nucleon
momentum is more approximate in $^3$He than in deuterium. Since one has a three-body final state
instead of a two-body final state one has to assume that the two spectator nucleons have no relative
momentum. To overcome this problem the recoil nucleon momentum has been calculated for nucleons
detected in TAPS using time-of-flight instead
of kinematics, which results in the cross sections in Fig.\ref{fig:qfeta} (right).  In this case
the
structure has a width of $45$ MeV which is comparable to
the experimental resolution.
\section{Conclusions}
The coherent $\eta$-photoproduction off $^3$He was measured with improved statistical quality. 
The resulting total cross section rises extremely between the coherent and breakup threshold. The
angular distributions at threshold (not shown) are almost isotropic or have
even an angular dependence opposite to the expectation from the form factor behavior. 
All these effects are strong evidence for dominant final state interaction, which could be related
to a resonant state at $\eta$-photoproduction threshold. \newline
The cross section of quasi-free $\eta$-photoproduction on the neutron shows a bump-like structure.
The
position and width of this structure is consistent with the deuteron data. The existence of this
structure in the cross section on $^{3}$He, which has a different neutron-to-proton ratio and a
bigger Fermi motion than deuterium, makes it very unlikely that this structure
is caused by rescattering of mesons or final state interaction. Currently further
experiments are running at MAMI in Mainz and ELSA in Bonn to measure single and double
polarisation observables which then can be used to identify the responsible partial waves.

\acknowledgements{%
This work was supported by Schweizerischer Nationalfonds, DFG, and EU/FP6. The results on the
$\eta$-mesic nuclei are part of the PhD-Thesis of F. Pheron.
}

%
}  


\end{document}